\documentclass[twocolumn,showpacs,preprintnumbers,amsmath,amssymb,prb]{revtex4-1}
\usepackage{graphicx}
\usepackage{dcolumn}
\usepackage{bm}
\usepackage{float}

\begin{document}

\title{Lindhard and RPA susceptibility computations in extended momentum space in electron doped 
cuprates}
\author{Yung Jui Wang$^1$, B. Barbiellini$^1$, Hsin Lin$^1$, 
Tanmoy Das$^1$, Susmita Basak$^1$, 
P.E. Mijnarends$^{1,2}$, S. Kaprzyk$^{1,3}$, R.S. Markiewicz$^1$ and 
A. Bansil$^1$}
\affiliation{
$^1$Physics Department, Northeastern University, Boston, Massachusetts
02115, USA\\
$^2$Department of Radiation, Radionuclides \& Reactors, Faculty of
Applied Sciences, Delft University of Technology, Delft, The
Netherlands \\
$^3$AGH University of Science and Technology, 30059 Krak\'{o}w, Poland\\
}

\date{\today}

\begin{abstract}
We present an approximation for efficient calculation 
of the Lindhard susceptibility $\chi^{L}(\textbf{q},\omega)$ in a
periodic system through the use of simple products of real 
space functions and the fast Fourier transform (FFT). 
The method is illustrated by providing $\chi^{L}(\textbf{q},\omega)$ results for 
the electron doped cuprate Nd$_{2-x}$Ce$_{x}$CuO$_{4}$ extended over several 
Brillouin zones. These results are relevant for interpreting inelastic X-ray 
scattering spectra from cuprates.
\end{abstract}

\maketitle
\def\thesection{\arabic{section}}

\section{introduction}
The dynamic structure factor $S(\textbf{q},\omega)$ is a useful function
of momentum and energy 
introduced by Leon Van Hove \cite{vanH}, which
contains information about density-density correlations and their time evolution. 
Experimentally, $S(\textbf{q},\omega)$ can be accessed most directly by Inelastic X-ray Scattering (IXS), 
which has acquired greater importance with the advent of powerful synchrotron sources \cite{PT}.
However, since X-rays are strongly absorbed in materials with high density, 
IXS may be suitable mostly for low-$Z$ systems. 
Nevertheless, in the case of heavier elements, recent studies have shown that 
if the photon incident energy is near an X-ray 
absorption edge the cross section can be enhanced, and 
the resulting resonant inelastic X-ray scattering 
(RIXS) offers a new window for probing both empty and filled electronic states 
\cite{schuelke,kotani,rueff}. 
Recent efforts to develop a 
first-principles formulation of the RIXS spectrum 
explore an interesting hypothesis
\cite{rixs1,rixs2,rixs3} that the 
RIXS cross-section is directly related to
$S(\textbf{q},\omega)$, complicating effects of 
the core hole notwithstanding. 
However, this claim remains controversial 
\cite{KAhn} and must be checked by 
testing the theory against accurate 
experimental results \cite{grenier}.
Cu-K-edge RIXS for cuprates \cite{schuelke,kim,Ishii,Collart}
probes the spectrum throughout momentum space encompassing many Brillouin zones.
Therefore an important theoretical task is to produce realistic calculations of the dynamic structure factor within the
framework of either Many-Body Perturbation Theory (MBPT) or Time-Dependent Density Functional Theory (TDDFT)
\cite{review} starting from a Lindhard susceptibility representing the response of an unperturbed Kohn-Sham system.
In particular, local field effects \cite{Adler,fleszar1,fleszar2,weiku} are known
to modify the spectral weight of both collective and single-particle excitations in the dynamic structure factor of solids.

In this study we focus on an approximation to efficiently calculate from one particle spectral functions 
the Lindhard susceptibility $\chi^{L}(\textbf{q},\omega)$. 
This approximation has successfully described the susceptibility of heavy rare earth elements \cite{natIDH}
and can also reliably describe the X-ray inelastic scattering momentum dependency 
in higher Brillouin zones
for an energy transfer $\omega$ where the single-particle excitations dominate.
As an example we consider paramagnetic
Nd$_{2-x}$Ce$_{x}$CuO$_{4}$ (NCCO), which has
a relatively simple, nearly two-dimensional metallic Cu-O band near the Fermi level. 
We identify important features throughout energy-momentum space and we delineate the 
specific manner in which $\mbox{Im}\chi^{L}(\textbf{q},\omega)$ decays
as a function of $\textbf{q}$. 
These results enable an assessment of the extent
to which $S(\textbf{q},\omega)$ reproduces the RIXS cross-section 
in a cuprate via direct comparison of the theory with corresponding experiments
in extended regions of the momentum space.

An outline of this paper is as follows. 
In Section 2, we 
present the relevant formalism.
The details of the
electronic structure methods and the numerical schemes 
are given in Section 3. 
The theoretical results for $\mbox{Im}\chi^{L}(\textbf{q},\omega)$ 
are presented
and discussed in Section 4, and the
conclusions are summarized in Section 5.

\section{Formalism}
In a periodic solid, 
the susceptibility becomes \cite{Adler} a tensor 
in the reciprocal lattice vector space {${\bf G}$}.
The fluctuation-dissipation theorem relates
the dynamical structure factor $S({\bf q},\omega)$ to
the susceptibility via
\begin{equation}
S(\textbf{q},\omega) = 
-\frac{1}{\pi}\sum_{\bf k,G}
\frac{\mbox{Im}[\chi_{{\bf G},{\bf G}}(\textbf{k},\omega)]}
{1-e^{-\hbar\omega/kT}}
\delta(\textbf{q}-\textbf{k}-{\bf G}).
\label{eqS}
\end{equation}
Thus, IXS experiments do not probe all matrix elements of the response 
$\chi_{\bf G,G'}({\bf k},\omega )$, but only the 
diagonal elements $\chi_{\bf G,G}({\bf k},\omega )$ \cite{abba}. 
If we approximate the susceptibility by the bare susceptibility
$\chi^{0}_{\bf G,G'}({\bf k},\omega )$ then 
\cite{fleszar1,fleszar2,weiku,KSturm}
\begin{align}
\mbox{Im}[\chi^0_{{\bf G},{\bf G}}(\textbf{k},\omega)]=
-\sum_{\nu,\mu}|\mbox{M}_{\bf G}^{\nu,\mu}|^2
\int_{-\omega}^{0} d\epsilon~
A_{\nu}(\epsilon) A_{\mu}(\epsilon+\omega).
\label{eqchi0a}
\end{align}
The matrix elements $\mbox{M}_{\bf G}^{\nu,\mu}$ 
can be expressed in the Dyson orbital basis set 
$g_{\nu}$ as \cite{kaplan,bba}
\begin{equation}
\mbox{M}_{\bf G}^{\nu,\mu}= 
<g_{\nu} |e^{i(\textbf{k}+\textbf{G})\textbf{r}} | g_{\mu}>
\label{eqchi0}.
\end{equation}
The spectral functions associated with the
Dyson orbitals are
\begin{equation}
A(\textbf{p},\omega)=\sum_{\nu} |g_{\nu}(\textbf{p})|^2
A_{\nu}(\epsilon),
\label{eqdyson1}
\end{equation} 
and
\begin{equation}
A_{\nu}(\omega)=
\frac{\gamma}{\pi[(\hbar\omega-\epsilon_{\nu})^{2}+\gamma^{2}]},
\label{eqA}
\end{equation}
where $\epsilon_{\nu}$ is the excitation energy
associated with the Dyson orbital $g_{\nu}$
and $\gamma$ is infinitesimally small
(see also Appendix A).
The Dyson orbitals can often be approximated reasonably  
by Bloch orbitals as
\begin{equation}
g_{\textbf{k},n}(\textbf{r})=\exp(i\textbf{k}\cdot\textbf{r})
\sum_{\textbf{G}} C_{\textbf{G}}^{\textbf{k},n}\exp(-i\textbf{G}\cdot\textbf{r}),
\label{eqbloch1}
\end{equation}
with the momentum density given by
\begin{equation}
|g_{\textbf{k},n}(\textbf{p})|^2=
\sum_{\textbf{G}}\delta(\textbf{p-k+G})~
|C_{\textbf{G}}^{\textbf{k},n}|^2.
\label{eqbloch2}
\end{equation}
The label $\nu=({\textbf{k},n})$ is a composite index that codes the Bloch
wave vector $\textbf{k}$ and the energy band index $n$.
The Fourier coefficients $C_{\textbf{G}}^{\textbf{k},n}$
of the periodic part of the Bloch function
are labeled by the reciprocal vectors $\textbf{G}$. 
In this case, the dynamical structure factor at $T=0$
becomes
\begin{align}
S({\bf q}, \omega) &= -\sum_{n,m,{\bf k},{\bf k}',
{\bf G}', {\bf G}'',{\bf G}} 
(C_{{\bf G}'}^{{\bf k},n})^* C_{{\bf G}' + {\bf G}}^{{\bf k}',m} 
C_{{\bf G}''}^{{\bf k},n} (C_{{\bf G}'' + {\bf G}}
^{{\bf k}',m})^* ~ \\ \nonumber
\times &\delta({\bf q}+{\bf k} - {\bf k}' - {\bf G}) 
\int_{-\omega}^0 d\varepsilon A_n({\bf k}, \varepsilon) A_m({\bf k}',
\varepsilon + \omega).
\label{eqfles}
\end{align}
The dominant part of $S({\bf q}, \omega)$ 
is given by the partial sum of the real positive terms
${\bf G}'={\bf G}''$.
Next, following Wen\cite{Wen176},
we neglect the remaining 
complex terms because the randomness of their 
phases produces destructive 
interferences. A few straightforward 
algebraic simplifications then yield
\cite{ng,schuelke}
\begin{equation}
\mbox{Im}\chi^{L}(\textbf{q},\omega)=
-\int_{-\omega}^{0} \frac{d\epsilon}{2\pi} 
\int\frac{d^{3}p}{(2\pi)^{3}} 
A(\textbf{p},\epsilon)
A(\textbf{p}+\textbf{q},\epsilon+\omega).
\label{eqsqw}
\end{equation}
Thus, our approximation scheme
leads to an expression for $\mbox{Im}\chi^{L}(\textbf{q},\omega)$ 
similar to the free fermion form \cite{Mahan}
but with the spectral function
$A(\textbf{p},\epsilon)$ expressed in terms of
the Bloch wave functions
instead of plane waves. 
The approximation of Eq.~(9) becomes exact 
when $q$ is large (see e.g. Ref. [27]).
As already noted above, the asymptotic
decay of the imaginary part of Lindhard susceptibility
as a function of $\textbf{q}$ is well
described within the present framework.
When $\textbf{q}$ is small, the 
most significant features
of the susceptibility are produced 
by band structure effects,
which are fully included in our approach.
Notably, the origin of major peaks
in the imaginary part of the susceptibility 
lies in FS nesting.\cite{MLSB,utfeld}
Therefore, we expect our scheme to produce a reasonable approximation
to the dynamical structure factor in materials.

\section{Computational Methods}
%
%
The Dyson orbitals $g_{\nu}$ needed for the calculation
of the spectral function $A(\textbf{p},\epsilon)$, as already noted, 
can be reasonably replaced by the Kohn Sham orbitals
obtained within the Density Functional Theory 
(DFT) \cite{bba}. For this purpose, 
the DFT band structure calculations in NCCO 
were performed within the Local Density
Approximation (LDA) using an all-electron,  
fully charge self-consistent
semi-relativistic (KKR) method \cite{ABkkr}. 
The crystal structure used for NCCO
was body centered tetragonal (space-group I4/mmm) with
lattice parameters given 
by Massidda {\em et al.} \cite{massidda}. 
A self-consistent solution was obtained for $x=0$ 
with a convergence of   the crystal potential 
to about $10^{-4}$ Ry.

To demonstrate our approach in a relatively simple but interesting case, we restrict the calculation to a single band, namely  
the copper-oxygen band near the Fermi level in NCCO.  
In particular, the possible contribution of the Nd f-electrons is neglected 
by removing the f orbital from the basis set after the Nd self-consistent potential has been obtained.\cite{foot1}
The electronic structure shown in Fig.~1(a) has been produced with the minority spin part of the self-consistent ferromagnetic potential.
The doping effects were treated within a rigid band model 
by shifting the Fermi energy to accommodate
the proper number \textit{x} of electrons \cite{armitage,Abfoot1,new1,new2}.
In the electron momentum density (EMD) calculations (see Ref. \onlinecite{comptonAB} for details), the
momentum mesh was given by a momentum step 
$(\Delta p_x, \Delta p_y, \Delta p_z)$ 
= $(1/128a,1/128a,1/2c) /(2\pi)$.\cite{Abfoot2,new3,new4}
The total number of momentum points is
$1.54\times10^{8}$ $\textbf{p}$ 
within a sphere of radius $17.6$ a.u.

\begin{figure}
\includegraphics[width=8.0cm]{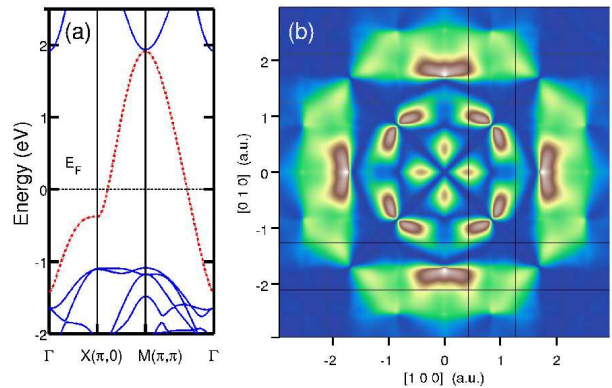}
\caption{(Color online)
(a) Band structure of NCCO near the Fermi level.
The CuO$_2$ band is shown by the red dotted line.
(b) Calculated integrated spectral function
$A(\textbf{p})$ for an isolated CuO$_2$ layer in NCCO. The zone
boundaries (for a simple tetragonal approximation) are marked by black lines.
Whites denote large values of $A(\textbf{p})$,
blues small values. The $A(\textbf{p})$ shown here contain
contributions only from the CuO$_2$ band.
}
\label{f1}
\end{figure}

We show in Fig.~\ref{f1}(a) 
the calculated band structure 
of NCCO near the Fermi 
level. The band closest to the Fermi level 
is shown 
by the red dotted curve and is
well isolated from other bands. 
This band ranges from $-1.4$ eV to $1.9$ eV
and the integral of the spectral function 
in this energy interval, evaluated with an energy resolution of  
$40$ meV, is   shown
in Fig.~\ref{f1}(b).
The two-dimensional spectral function $A(\textbf{p},\omega)$ 
is calculated by neglecting the 
small $k_{z}$ dispersion in the three dimensional electronic band structure 
\cite{rsbob}.
Similar EMD results for NCCO 
have been obtained within the LMTO\cite{blandin,bdj}. 
The resulting momentum density 
has the same symmetry as the 
copper-oxygen $d_{x^{2}-y^{2}}-p_{x,y}$
states in real space which form this energy band 
since the wave function in momentum space is the Fourier transform 
of the wave function in real space.
Fig.~\ref{f1}(b) shows that the low intensity along the $x-y$ diagonal direction in the 2D-EMD map is a signature of $d_{x^{2}-y^{2}}$ symmetry. 
Moreover, since the radial momentum dependence of an atomic state
of angular momentum $\ell$ behaves as $p^{\ell}$ 
at small momenta \cite{Mijnarends:1973}, 
the 2D-EMD intensity at low momenta is from the O-$2p$ orbitals,  
while the Cu-$3d_{x^{2}-y^{2}}$ orbitals contribute at higher momenta.\cite{Abfoot3,new5}
This implies that the signal coming from the O-2$p$ states is more visible in the first 
Brillouin Zone while the Cu-3$d$ states are better seen in higher Brillouin zones.
We can see from Fig.~\ref{f1}(b) that the 2D-EMD intensity is strongly modulated by wave function effects, which suggests that the behavior of 
$\mbox{Im}\chi^{L}(\textbf{q},\omega)$ in NCCO in different zones will also be modified by 
these effects; 
however, our approximation
in Eq.~(9) neglects some interference effects produced by the phases of the Fourier coefficients of the Bloch
wave functions.

Equation \ref{eqsqw} shows that the $\mbox{Im}\chi^{L}(\textbf{q},\omega)$ at zero temperature can be 
written as a convolution of two spectral functions.  
This $\mbox{Im}\chi^{L}(\textbf{q},\omega)$ captures electron-hole excitations described by 
Dyson orbitals \cite{kaplan,bba} but does not include collective excitations such as 
plasmons or phonons. For efficient $\mbox{Im}\chi^{L}(\textbf{q},\omega)$ calculations, we replace the momentum space convolution of 
$A(\textbf{p},\omega)$ by a simple product of spectral functions
$B(\textbf{r},\omega)$ in real space given by 
\begin{equation}
B(\textbf{r},\omega)=\int\frac{d^{3}p}{(2\pi)^{3}} A(\textbf{p},\omega)
\exp(i\textbf{p~r}) ~.
\end{equation}
This enables us to take advantage of the fast Fourier transform (FFT) efficiency using the
convolution theorem \cite{rojas}.
The advantage of our FFT based method can be seen by comparing the computation time of
the FFT method with the time needed to directly compute $\mbox{Im}\chi^{L}(\textbf{q},\omega)$
via Eq.~\ref{eqsqw} using two matrices of size $2049\times2049$. The CPU time for 
the FFT method is $12$ seconds, while the direct computation takes $24$ minutes on the 
same machine\cite{ft_cpu}.

\begin{figure}
\includegraphics[width=8.5cm]{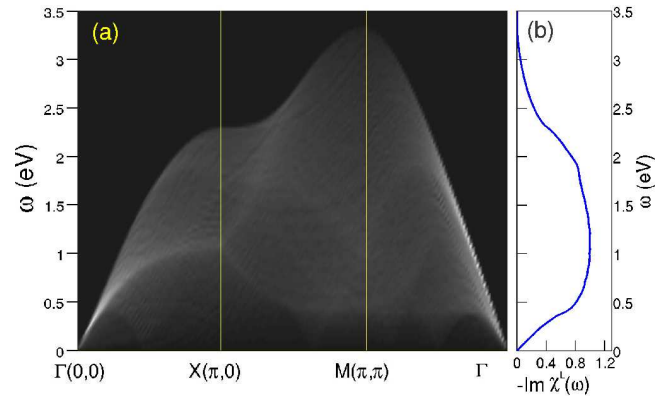}
\caption{(Color online)
(a) Calculated $-\mbox{Im}\chi^{L}(\textbf{q},\omega)$ of NCCO along high symmetry lines as a function of transition
energy. Whites denote the largest $-\mbox{Im}\chi^{L}(\textbf{q},\omega)$, 
blacks the smallest. 
(b) The integrated value of 
$-\mbox{Im}\chi^{L}(\textbf{q},\omega)$ over $\textbf{q}$ vs transition energy $\omega$.
}
\label{f2}
\end{figure}
\begin{figure}
\includegraphics[width=7cm]{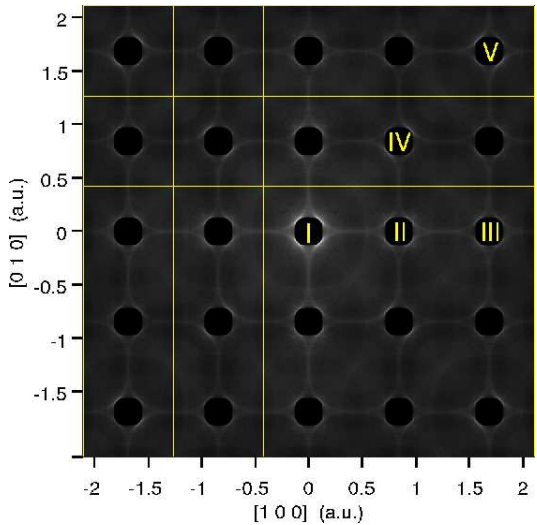}
\caption{(Color online)
Calculated $-\mbox{Im}\chi^{L}(\textbf{q},\omega)$ of NCCO at $\omega = 1.04$ eV plotted over several Brillouin 
zones. Black circles near the $\Gamma$-points indicate regions of zero intensity, 
as in Fig.~2. 
Index \MakeUppercase{\romannumeral 1} labels the first zone, indices \MakeUppercase{\romannumeral 2}, 
\MakeUppercase{\romannumeral 3} mark zones along 
the $(\pi,0)$ direction, and \MakeUppercase{\romannumeral 4} and \MakeUppercase{\romannumeral 5} 
the zones along the $(\pi,\pi)$ direction. 
}
\label{f3}
\end{figure}
\begin{figure}
\includegraphics[width=7cm]{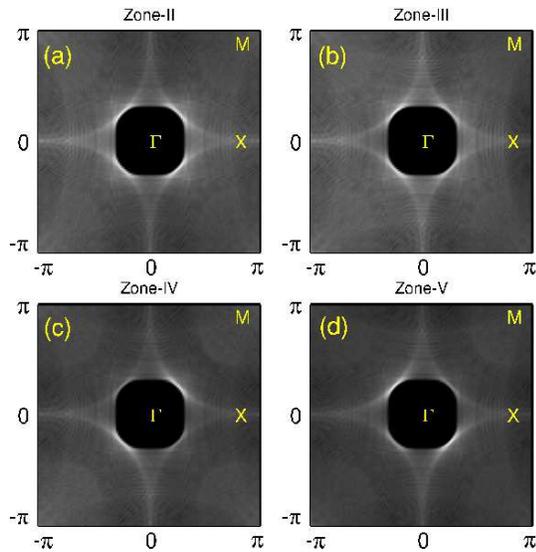}
\caption{(Color online)
(a-d) Contour maps of $-\mbox{Im}\chi^{L}(\textbf{q},\omega = 1.04$ eV) of NCCO
in four different Brillouin zones. The symmetry of the maps in zone II and zone III
is very similar.
}
\label{f4}
\end{figure}
\section{Results}
We discuss our results with reference to Figures~2-5. 
In Fig.~\ref{f2}{a}, 
we show $-\mbox{Im}\chi^{L}(\textbf{q},\omega)$ along high symmetry 
lines as a function of $\omega$. 
The black part of this figure marks the region of zero intensity
where no electron hole transitions are available. Strong 
intensity seen near $1$ eV around $(\pi,0)$ is due to 
a sort of a Van Hove singularity in $\mbox{Im}\chi^{L}(\textbf{q},\omega)$, which 
is associated with the high energy kink or the waterfall effect in the electronic spectrum\cite{waterfall}. 
When we compare our Fig.~\ref{f2}{a} to the experimental RIXS spectrum of overdoped NCCO presented in 
Ref. \onlinecite{ywli}, 
we find that the experiment is well described 
by the $\textbf{k}$ resolved joint density of states despite the complicating effects of the core-hole\cite{rixs3}. 
In particular, the features in the lowest experimental RIXS band within 
the energy range of $\omega=0.5$ to $\omega=2$eV are well reproduced by our calculations. The integrated value of 
$-\mbox{Im}\chi^{L}(\textbf{q},\omega)$ over $\textbf{q}$, plotted in Fig.~\ref{f2}{b}, yields the total number of electron-hole transitions at a 
given energy. Since the highest peak in Fig.~\ref{f2}{b}
is located at $1.04$ eV, we focus on analyzing $-\mbox{Im}\chi^{L}(\textbf{q},\omega)$ at this 
particular energy in the remainder of this article.
$-\mbox{Im}\chi^{L}(\textbf{q},\omega)$ is shown in Fig.~\ref{f3}
for $\omega=1.04$ eV over several Brillouin zones 
marked by yellow lines. The first Brillouin zone, 
located at the center of the figure, has the highest intensity.
The intensity is seen to decrease slowly as $q$ increases, and interesting patterns due to d electron wavefunction effects appear in higher zones.
In the first zone, Fig.~\ref{f3}, some strong peaks are present surrounding 
the zero-intensity hole centered at $\Gamma$ with a relatively low intensity 
appearing at the zone corners $M$.

\begin{figure}
\includegraphics[width=7cm]{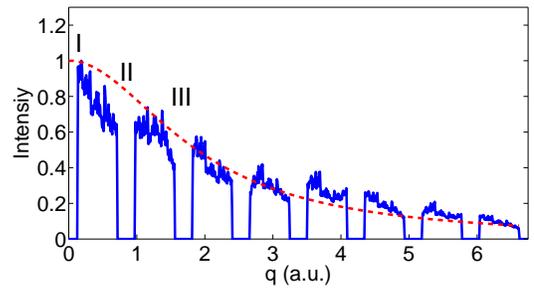}
\caption{(Color online)
Cuts of $-\mbox{Im}\chi^{L}(\textbf{q},\omega = 1.04$ eV) taken along $(\pi,0)$ in Fig.~3 (extended to higher BZs).
The highest intensity has been normalized to one. The zone-to-zone change in 
$-\mbox{Im}\chi^{L}(\textbf{q},\omega = 1.04$ eV) is mainly an overall decrease in the intensity with smaller changes due to 
matrix element effects. 
The envelope of $-\mbox{Im}\chi^{L}(\textbf{q},\omega)$ is fit by using a simple Lorentzian shape
given by the red dashed line.
Labels I, II and III correspond to 
the zone indices in Fig.~3.}
\label{f5}
\end{figure}

Further details of $-\mbox{Im}\chi^{L}(\textbf{q},\omega)$ are shown in Figs.~\ref{f4}(a)-(d),  which are
blow ups of the four Brillouin zones marked by \MakeUppercase{\romannumeral 2}, 
\MakeUppercase{\romannumeral 3},
\MakeUppercase{\romannumeral 4}, and \MakeUppercase{\romannumeral 5} in Fig.~\ref{f3}.
The Brillouin zones displayed in 
Figs.~\ref{f4}(a)-(d) show a similar 
overall pattern but modulated with subtle matrix element effects.
For instance, regions of strong intensity spread towards $(-\pi,0)$ in zone 
\MakeUppercase{\romannumeral 2} (Fig.~\ref{f4}{a}),
but towards $(-\pi,\pm\pi)$ in zone \MakeUppercase{\romannumeral 3} (Fig.~\ref{f4}{b}). The 
intense (bright) peaks point along 
one diagonal direction 
in zone \MakeUppercase{\romannumeral 4} in Fig.~\ref{f4}{c}, 
but are rotated by 90 degrees in zone \MakeUppercase{\romannumeral 5} in Fig.~\ref{f4}{d}. 
Figure~\ref{f5} presents a cut through $-\mbox{Im}\chi^{L}(\textbf{q},\omega)$ along the [100] direction 
in order to illustrate the decay of $-\mbox{Im}\chi^{L}(\textbf{q},\omega)$ 
as a function of momentum transfer $q$.
The highest intensity has been normalized to unity for ease of comparison. 
Surprisingly, at momenta as large as 6 a.u. one can still see
features with amplitude exceeding 10 \% of the highest intensity
(located in the first Brillouin zone). This effect can be explained
by the fact that d electron particle-hole transitions
can involve particularly high momentum transfers.
We can fit the envelope of $-\mbox{Im}\chi^{L}(\textbf{q},\omega)$ by using a simple Lorentzian shape
$\frac{1}{1+(\frac{q}{k_0})^{2}}$
with $k_0 = 1.89$ a.u.\cite{footnotefsum} 
Since RIXS has often been thought to be related to $S(q,\omega )$, it is an interesting question 
whether a similar decay factor $k_0$ is found in RIXS experiments.  Our results thus provide a new way to 
test the hypothesis that the RIXS cross-section is directly related to $S(\textbf{q},\omega)$. 

\begin{figure}
\includegraphics[width=7cm]{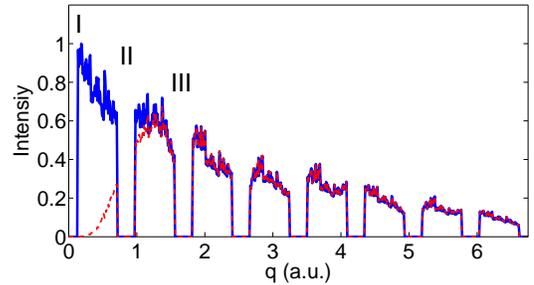}
\caption{(Color online)
Cuts of $-\mbox{Im}\chi^{RPA}(\textbf{q},\omega = 1.04$ eV) (in red dashed line)  
and $-\mbox{Im}\chi^{L}(\textbf{q},\omega)$ (in blue solid line)   
taken along $(\pi,0)$ in Fig.~3 (extended to higher BZs).
The highest intensity of each curve has been scaled by taking the highest intensity of $-\mbox{Im}\chi^{L}(\textbf{q},\omega)$ as unity.
Labels I, II and III correspond to 
the zone indices in Fig.~3. No plasmon peak is present in 
this energy slice $\omega = 1.04$ eV. }
\label{f6}
\end{figure}

We have also obtained the real part of Linhard susceptibility $\chi^{L}$ 
by applying Kramers-Kronig relation.
However, the Linhard susceptibility gives only the response of the independent
electrons to the external potential. In order to estimate the effect of 
screening effects, one can consider the susceptibility within 
the random phase approximation (RPA) given by
$\chi^{RPA}=\chi^{L}/(1-V_{q}\chi^{L})$, where $V$ is the Coulomb
interaction decaying as $\sim 1/q^{2}$. 
In this approach, the sharp singularities of $\chi^{RPA}$ 
due to the denominator give the plasmon modes.
Figure 6 illustrates the corrections are more important when
the external perturbation is of very long wavelength (i.e. $q$ small)\cite{footnotelucia}.
Interestingly when $q$ is of the order of $k_0$,
$-\mbox{Im}\chi^{RPA}(\textbf{q},\omega)$ 
recovers back to the $-\mbox{Im}\chi^{L}(\textbf{q},\omega)$.

\section{conclusions}
We have presented a formalism for a first principles computation 
of the Lindhard susceptibility 
$\chi^{L}(\textbf{q},\omega)$ in extended momentum space. 
We have demonstrated a tremendous improvement in performance 
by calculating
$\mbox{Im}\chi^{L}(\textbf{q},\omega)$ through an approximation involving
products of real space spectral functions 
$B(\textbf{r},\omega)$
and FFTs instead of using the standard approach
involving costly matrix multiplications.
Our theoretical $\mbox{Im}\chi^{L}(\textbf{q},\omega)$ results for the doped cuprate NCCO 
will allow a detailed comparison with the RIXS experiments, and hence an assessment
of the extent to which  $\mbox{Im}\chi^{L}(\textbf{q},\omega)$ represents 
a good approximation to the RIXS cross section. 
The present work also provides a realistic linear response based 
starting point for developing a many-body 
perturbation theory of particle-hole excitations 
within the DFT framework. 
%

We are grateful to J. Lorenzana for discussions.
This work is supported by the US Department of Energy, Office of
Science, Basic Energy Sciences contracts DE-FG02-07ER46352 and DE-SC0007091 (CMCSN), and benefited 
from the allocation of supercomputer time at NERSC and 
Northeastern University's Advanced Scientific Computation Center (ASCC).
It was also sponsored by the Stichting Nationale Computer Faciliteiten (NCF)
for the use of supercomputer facilities, with financial support from NWO
(Netherlands Organization for Scientific Research).

\appendix
\section{Relation between susceptibility and spectral function}
We introduce the susceptibility matrix element 
\begin{align}
\mbox{F}^{\nu,\mu}=
\frac{f(\epsilon_{\nu})-f(\epsilon_{\mu})}
{\hbar\omega+
\epsilon_{\nu} - \epsilon_{\mu}
+i\gamma}.
\label{eq:3}
\end{align}
where $f(\epsilon)$ is the Fermi function.
The term 
$\mbox{Im}[\mbox{F}^{\nu,\mu}]$ 
can be also written 
in terms of the spectral function $A_{\nu}$.
By using
\begin{equation}
A_{\nu}(\omega)=
\mbox{Im}[\frac{1}{\hbar\omega-\epsilon_{\nu}+i\gamma}]~,
\label{eq:14}
\end{equation}
we obtain
\begin{equation}
\mbox{Im}[\mbox{F}^{\nu,\mu}]=
-\int_{-\omega}^{0} d\epsilon~ 
A_{\nu}(\epsilon) A_{\mu}(\epsilon+\omega)~.
\label{eq:20}
\end{equation}


\begin{thebibliography}{99}
\bibitem{Abfoot1} The rigid band model is expected to be a good approximation for doping away from the cuprate layers. 
It will however be interesting to examine doping effects by using
supercell or other first principles approaches.\cite{new1,new2}
\bibitem{Abfoot2} EMD can be probed directly via Compton scattering\cite{new3} or positron-annihilation\cite{new4} experiments.
\bibitem{Abfoot3} Similar matrix element effects are seen in angle-resolved photoemission spectra of the cuprates.\cite{new5}
\bibitem{ft_cpu}
The CPU used in the calculations is the Intel Core i7-920.
\end{thebibliography}
\end{document}